\documentclass[sigconf,10pt]{acmart}

\usepackage{xspace}
\usepackage{enumitem}
\usepackage{subfigure}
\usepackage{subcaption}
\usepackage{caption} 
\usepackage[normalem]{ulem}
\usepackage{graphicx} 
\usepackage{multirow}
\usepackage{makecell}
\usepackage{float}
\usepackage{siunitx}
\usepackage{fp}
\usepackage{algorithm}
\usepackage{algpseudocode}

\newcommand{\SystemName}{\textsc{VoxAnchor}\xspace}
\definecolor{mybarcolor}{HTML}{bec7e2}
\newlength\maxlentime
\newcommand\timebar[3][mybarcolor]{%
  \FPeval\raw{round((#3-1.5)/#2:4)}%
  \FPeval\result{max(\raw,0.015)}%
  \rlap{\textcolor{#1}{\hspace*{\dimexpr-\tabcolsep+.5\arrayrulewidth}%
        \rule[-.05\ht\strutbox]{\result\maxlentime}{.95\ht\strutbox}}}%
  \makebox[\dimexpr\maxlentime-0.2\tabcolsep+\arrayrulewidth][r]{~#3}}
\def\headertime{New}
\renewcommand\footnotetextcopyrightpermission[1]{} 
\begin{document}

\title{VoxAnchor: Grounding Speech Authenticity in Throat Vibration via mmWave Radar}

\author{Mingda Han$^{1}$, Huanqi Yang$^{2}$, Chaoqun Li$^{1}$, Wenhao Li$^{2}$, Guoming Zhang$^{1}$,\\ Yanni Yang$^{1}$, Yetong Cao$^{1}$, Weitao Xu$^{2}$, Pengfei Hu$^{1}$}
\affiliation{
\institution{$^1$Shandong University, $^2$City University of Hong Kong
}
\country{}
}
\renewcommand{\shortauthors}{Han et al.}

\begin{abstract}
Rapid advances in speech synthesis and audio editing have made realistic forgeries increasingly accessible, yet existing detection methods remain vulnerable to tampering or depend on visual/wearable sensors.
In this paper, we present \SystemName, a system that physically grounds audio authentication in vocal dynamics by leveraging the inherent coherence between speech acoustics and radar-sensed throat vibrations.
\SystemName uses contactless millimeter-wave radar to capture fine-grained throat vibrations that are tightly coupled with human speech production, establishing a hard-to-forge anchor rooted in human physiology.
The design comprises three main components: 
(1) a cross-modal framework that uses modality-specific encoders and contrastive learning to detect subtle mismatches at word granularity; 
(2) a phase-aware pipeline that extracts physically consistent, temporally faithful throat vibrations; and 
(3) a dual-stage strategy that combines signal-level onset detection and semantic-level coherence to align asynchronous radar and audio streams. 
Unlike liveness detection~\cite{zhang2016voicelive}, which only confirms whether speech occurred, \SystemName verifies what was spoken through word-level content consistency, exposing localized edits that preserve identity and global authenticity cues.
Extensive evaluations show that \SystemName achieves robust, fine-grained detection across diverse forgeries (editing, splicing, replay, deepfake) and conditions, with an overall EER of 0.017, low latency, and modest computational cost.
\end{abstract}

\maketitle
    \vspace{-0.1in}
\section{Introduction}

Audio forgeries, ranging from surgically edited recordings to AI-generated deepfakes, are rapidly undermining the reliability and security of voice communication~\cite{khanjani2023audio}. 
Editing-based attacks can manipulate timing, semantics, or speaker identity without leaving obvious acoustic artifacts, while modern deepfake models synthesize highly realistic speech that is difficult to distinguish even for trained forensic analysts~\cite{casanova2022yourtts, kim2021conditional, popov2021grad}. 
Such capabilities have already precipitated real-world crises. 
For instance, manipulated audio was used to authorize a fraudulent \$35 million transfer~\cite{ClonedVoice}, and to mislead voters through impersonated political robocalls~\cite{guardian2024robocalls}.
These incidents exemplify how convincingly fake audio can be weaponized, eroding trust in voice as a reliable medium for communication and evidence. 
As such forgeries become increasingly indistinguishable from reality, establishing a robust mechanism to verify the authenticity and integrity of audio content is critical for maintaining public trust and institutional integrity.

\begin{figure}
    \centering
    \includegraphics[width=1\linewidth]{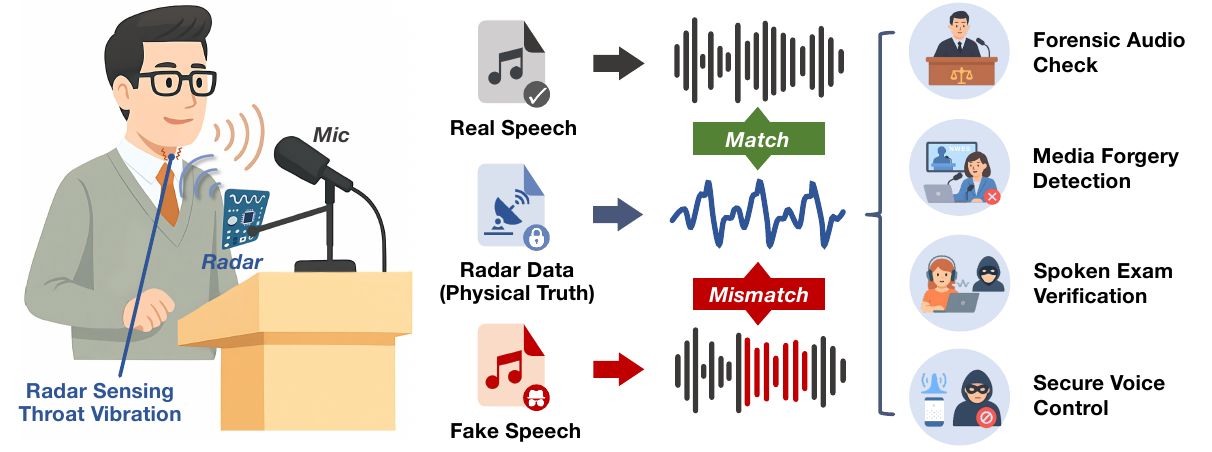}
    \vspace{-5mm}
    \caption{\SystemName anchors audio in radar-sensed throat vibrations to detect subtle forgeries via cross-modal consistency.}
    \label{fig:illustration}
    \vspace{-0.25in}
\end{figure}

Traditional audio forgery detection methods can be broadly categorized into two types: content-based and consistency-based. Content-based methods, including both classical machine learning~\cite{sahidullah2015comparison,javed2022voice} and modern deep learning models~\cite{gomez2019light,chitale2024hybrid,tak2021end}, focus on either handcrafted or deep-learned audio features.
Meanwhile, consistency-based approaches verify whether the audio aligns with auxiliary visual modalities. 
For example, some assess lip–speech consistency in video~\cite{bohacek2024lost,Zhou_2021_ICCV}, while others like TalkLock~\cite{shahid2023my} embed signed audio features into visual QR codes to enforce cross-modal alignment during recording. 
Though effective under certain conditions, content-based methods often miss manipulations such as splicing or segment-level edits that preserve natural acoustic statistics~\cite{li2025measuring}, and modality-consistency methods rely on video input, restricting them to face-to-camera scenarios.

To address these limitations, recent efforts have explored physically grounded solutions.
Blue et al.~\cite{blue2022you} reconstruct vocal-tract parameters from audio to detect synthesis artifacts, and VoiceRadar~\cite{kumari2025voiceradar} analyzes acoustic plausibility via room impulse responses, but both remain in the digital domain. Concurrently, OriginStory~\cite{originstory2025asu} proposes grounding media authenticity in human physiological signals.
F$^2$Key~\cite{duan2024f2key} captures facial acoustic reflections via earphones to derive cryptographic keys, and Aegis~\cite{gao2025exploring} uses co-modulated ultrasound for forgery detection, but both require user-worn accessories (earphones or ultrasound emitters), limiting deployment flexibility.

An ideal anti-forgery system should capture physical speech cues that are inherently linked to the audio and hard to forge, while remaining contact-free and non-intrusive at capture time.
Among such cues, throat vibrations encode rich physiological speech traits that are tightly coupled with vocal production and harder to mimic than lip motion~\cite{liu2023wavoid}, offering a stronger basis for cross-modal verification.
Meanwhile, recent advances in millimeter-wave (mmWave) sensing~\cite{hu2023mmecho,hu2025tracking,han2024mmsign} enable contactless detection of sub-millimeter vibrations, offering a practical and unobtrusive means to capture speech-related motion.
Building on these insights, we present \SystemName, a non-intrusive system that verifies speech authenticity by assessing cross-modal coherence between mmWave-sensed throat vibrations and audio.
While fake audio can imitate what is heard, it cannot recreate how speech is physically produced; \SystemName exploits this asymmetry by using vocal-induced throat vibration as a hard-to-forge anchor.
Crucially, \SystemName differs from both liveness detection and speaker verification: liveness detection only confirms that speech occurred, and speaker verification checks \emph{who} spoke, whereas \SystemName verifies \emph{what} was spoken by detecting word-level content inconsistencies between audio and its physiological origin.
As illustrated in Fig.~\ref{fig:illustration}, only genuine speech maintains strong cross-modal coherence rooted in shared physiology.

Despite its promise, implementing \SystemName involves three key challenges:
\textbf{1) Word-level sensitivity.} Although mmWave and audio signals share a common articulatory source, their low-level correlation is too coarse for subtle, word-level manipulations. As shown in Sec.~\ref{subsec:CrossModalCorrelation}, simple time- or frequency-domain metrics struggle with word-level forgeries. The core challenge is to design expressive cross-modal representations and comparison methods that reveal fine-grained inconsistencies. 
\textbf{2) Physically faithful vibration recovery.} Existing mmWave speech sensing methods typically rely on deep learning to reconstruct audio, optimizing for semantic similarity rather than physical consistency with throat vibrations. Consequently, the recovered features lack temporal continuity and physical interpretability. Therefore, the second challenge is to extract throat micro-vibrations with high physical fidelity as a trustworthy anchor for consistency analysis.
\textbf{3) Multi-level cross-modal alignment.} mmWave and audio are captured by separate devices without a shared clock. Due to their differing sensitivities and response characteristics, envelope-based methods often fail to achieve precise synchronization. Moreover, mmWave captures low-level articulations, while audio reflects higher-level semantics and prosody, leading to representational asymmetry. The third challenge lies in designing a unified alignment strategy that bridges both signal-level timing and semantic-level correspondence.

We address these challenges with an integrated approach combining signal processing, alignment, and multi-modal learning.
First, to enable word-level tampering detection, we build a cross-modal verification framework that captures semantic dependencies between mmWave and audio signals. By comparing their representations at word-level granularity, our system can identify subtle mismatches caused by fine-grained manipulations.
Second, we design a phase-aware signal processing pipeline to extract speech-induced throat micro-vibrations from radar echoes with high physical fidelity. Unlike existing learning-based approaches that focus on reconstructing intelligible speech, our method emphasizes physical interpretability and temporal continuity, providing a more reliable basis for consistency analysis.
Third, to tackle the alignment challenge, we propose a coarse-to-fine synchronization strategy. 
Coarse alignment is achieved via signal-level onset detection, then refined using a semantic-level coherence loss based on normalized cross-correlation, enabling robust synchronization under modality mismatch.
The contributions of this paper can be summarized as follows:
\vspace{-0.03in}
\begin{itemize}[leftmargin=*]
    \item We propose \SystemName, a system that grounds audio authentication in throat vibrations. It establishes a physically grounded paradigm of cross-modal forgery detection by exploiting radar-audio coherence to achieve fine-grained, word-level content verification beyond purely acoustic methods.
    
    \item We design an end-to-end framework that integrates physically interpretable vibration extraction, cross-modal synchronization at both signal and semantic levels, and contrastive semantic fusion, enabling trustworthy word-level audio forgery detection.

    \item Extensive experiments validate the effectiveness and robustness of our method against diverse forgeries and conditions. With low latency and modest computational cost, it shows strong potential for real-world applications.
\end{itemize}
\vspace{-0.03in}

\begin{figure}[t]
    \centering
    \includegraphics[width=1\linewidth]{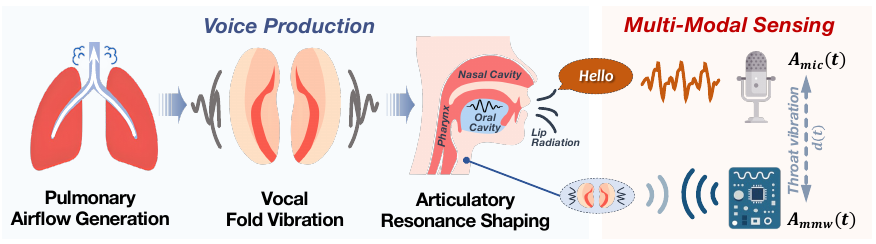}
    \vspace{-0.3in}
    \caption{Human voice production and multi-modal sensing.}
    \label{fig:preliminary}
    \vspace{-0.2in}
\end{figure}

\section{Preliminaries}
\label{sec:preliminary}

\subsection{Understanding Human Speech}
\label{sec:preliminary_speech}
Human speech production is a multifaceted process that transforms cognitive intentions into audible language through a series of coordinated stages.
As illustrated in Fig.~\ref{fig:preliminary}, the generation of human speech is a complex physiological process involving coordinated activity among the lungs, larynx, and upper vocal tract. 
This process aligns with the Source-Filter Theory of speech production~\cite{fant1971acoustic}, which models speech as a glottal source shaped by the vocal tract filter.

Airflow from the lungs initiates vocal fold vibration, producing a quasi-periodic glottal waveform $S_{\text{glottal}}(t)$ governed by the fundamental frequency and harmonics. 
Physically, vocal-fold motion can be represented by vertical displacement $d(t)$,
and the glottal source can be approximated by~\cite{cataldo2006synthesis}:
\begin{equation} \footnotesize
\setlength\abovedisplayskip{0.15cm}
\setlength\belowdisplayskip{0.15cm}
    S_{\text{glottal}}(t) = l \cdot \dot{d}(t)^2 + (A_0 + l \cdot d(t)) \cdot \ddot{d}(t),
\end{equation}
where $A_0$ is the initial glottal area, and $l$ is the glottal width.
The glottal signal is then filtered by the vocal tract, including the pharynx, oral, and nasal cavities, typically modeled as an impulse response $H_{\text{vt}}(t)$. 
It is then radiated through the lips $R_{\text{lip}}(t)$, forming the complete speech signal:
\begin{equation} \footnotesize \label{equ:audioGen}
\setlength\abovedisplayskip{0.15cm}
\setlength\belowdisplayskip{0.15cm}
    A(t) = S_{\text{glottal}}(t) * H_{\text{vt}}(t) * R_{\text{lip}}(t).
\end{equation}
This formulation captures the sequential transformations from source excitation to audible speech.
\vspace{-0.15in}

\subsection{Cross-Modal Correlation Analysis}
\label{subsec:CrossModalCorrelation}
We explore the correlation between mmWave-detected vocal vibrations and microphone-recorded speech, which is key to understanding audio authenticity.

\subsubsection{mmWave Signal}
mmWave radar captures speech by sensing micro throat movements during phonation, which induce phase shifts in the reflected continuous-wave signals.
We can extract this information from the micro-Doppler signatures in the radar echoes.
Thus, the mmWave signal modulated by vocal-induced motion can be expressed as
\begin{equation} 
\footnotesize
\setlength\abovedisplayskip{0.15cm}
\setlength\belowdisplayskip{0.1cm}
M(f,t) = H_0(f,t) + \sum_i A_i(f,t) e^{-j(\frac{2\pi}{\lambda} (l_{i0} + d_i(t)))},
\end{equation}
where $H_0(f,t)$ represents the static path, the second term models the dynamic paths caused by throat displacements $d_i(t)$ with initial path length $l_{i0}$, and $\lambda$ is the wavelength.

To extract vocal vibrations, we focus on the received signal phase. Assuming the static component $H_0(f,t)$ dominates the magnitude, the phase of $M(f,t)$ can be approximated as 
\begin{equation}
\footnotesize
\setlength\abovedisplayskip{0.15cm}
\setlength\belowdisplayskip{0.15cm}
    \phi(t) = \angle M(f,t) \approx -\frac{2\pi}{\lambda} d(t) + \varphi_0,
\end{equation}
where $d(t)$ is the effective throat displacement, and $\varphi_0$ is the static phase bias. This relationship between displacement and phase forms the basis of mmWave-based vibration sensing.

\subsubsection{Microphone Signal}
In contrast to radar, which directly senses sub-millimeter surface displacements of the skin, microphones capture speech by detecting pressure fluctuations in the air caused by the radiated sound wave. The signal $A(t)$ described in Eq.~\ref{equ:audioGen} is radiated from the lips into the surrounding air as an outgoing acoustic pressure wave $P$~\cite{mellow2012acoustics}:
\begin{equation} 
\footnotesize
\setlength\abovedisplayskip{0.15cm}
\setlength\belowdisplayskip{0.15cm}
    P(t, r) = \frac{Z_0}{4\pi r} \cdot \frac{d}{dt} A\left(t - \frac{r}{v}\right),
\end{equation}
where $r$ is the distance from the mouth to the microphone, $v$ is the speed of sound, and $Z_0$ is the acoustic impedance of air. This equation captures both the temporal delay and amplitude attenuation due to spatial propagation, as well as the angular dependence of the emitted sound.
As the pressure wave reaches the microphone, it introduces diaphragm vibrations proportional to the instantaneous pressure, generating the electrical signal $A_{\text{mic}}(t)$.
This signal reflects the compounded effects of glottal excitation, vocal tract filtering, lip radiation, and air propagation, resulting in a nonlinear mapping from the physiological vibration.

\begin{figure}
\centering
\subfigure[Sentence-level similarity.]{
    \begin{minipage}[t]{1\linewidth}
    \centering
    \includegraphics[width=1\linewidth]{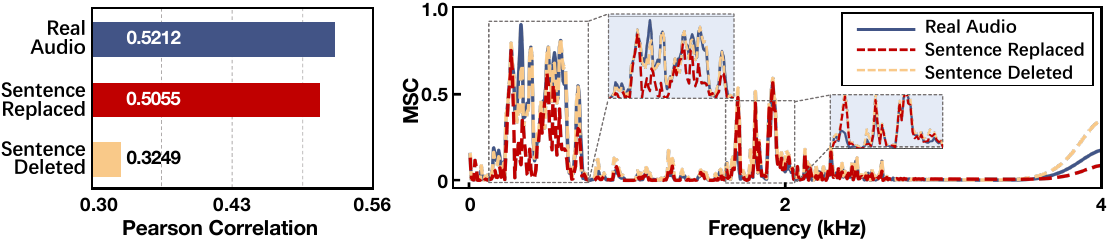}
    \vspace{-1.5em}
    \label{fig:msc1}
    \end{minipage}
}\vspace{-1em}
\subfigure[Word-level similarity.]{
    \begin{minipage}[t]{1\linewidth}
    \centering
    \includegraphics[width=1\linewidth]{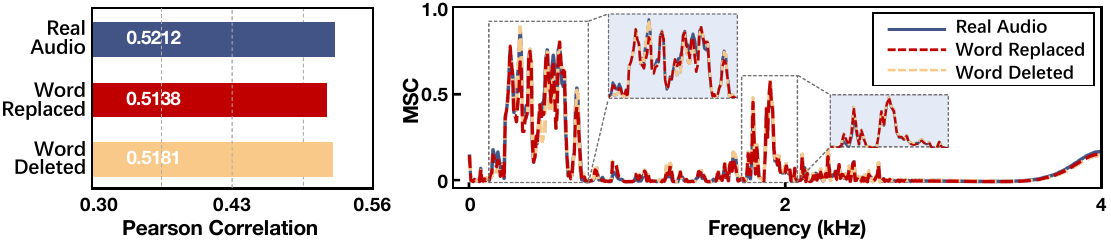}
    \vspace{-1.5em}
    \label{fig:msc2}
    \end{minipage}
}
\centering
\caption{Temporal-spectral correlation across forgery granularities.}
\label{fig:correlationAnalysis}
\vspace{-2em}
\end{figure}

\begin{figure*}
    \centering
    \includegraphics[width=0.95\linewidth]{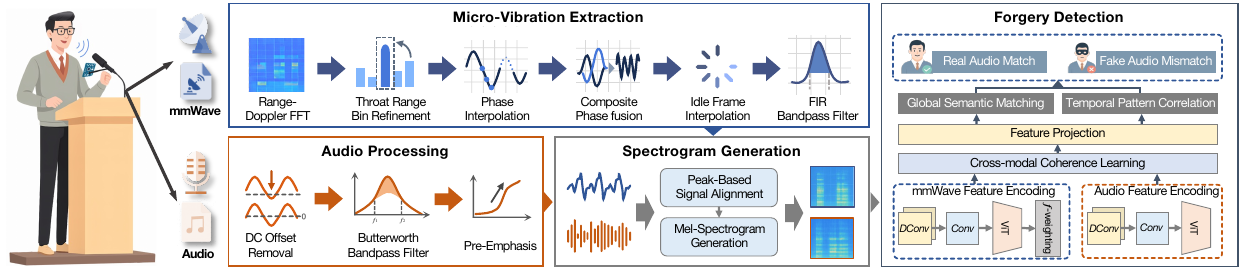}
    \vspace{-1em}
    \caption{Overall architecture of \SystemName. 
    }
    \label{fig:overview}
    \vspace{-1em}
\end{figure*}

\subsubsection{Correlation Analysis}
Although mmWave and microphone signals exhibit distinct representational forms, they both originate from the same underlying physical process, vocal fold and throat vibration $d(t)$, as illustrated in Fig.~\ref{fig:preliminary}.
This shared origin can be formalized as
\begin{equation}
\footnotesize
\setlength\abovedisplayskip{0.15cm}
\setlength\belowdisplayskip{0.15cm}
\begin{aligned}
& A_{\text{mmw}}(t) = \mathcal{F}_{\text{mmw}}\{d(t)\} + \varepsilon_m(t), \\
& A_{\text{mic}}(t) = \bigl[\mathcal{F}_{\text{src}}\{d(t)\} * H_{\text{vt}}(t) * R_{\text{lip}}(t)\bigr] + \varepsilon_a(t),
\end{aligned}
\end{equation}
where \(\mathcal{F}_{\text{mmw}}(\cdot)\) and \(\mathcal{F}_{\text{src}}(\cdot)\) map displacement to mmWave phase and glottal source. \(H_{\text{vt}}(t)\) and \(R_{\text{lip}}(t)\) are vocal-tract filtering and lip radiation, while \(\varepsilon_m(t)\) and \(\varepsilon_a(t)\) account for noise and unmodeled effects.
While mmWave radar provides a near-direct mapping from throat displacement to phase, microphone signals are shaped by multiple acoustic transformations that encode richer semantics. 
Despite lacking full semantic detail, mmWave signals preserve articulation patterns that are harder to spoof.

To assess this cross-modal consistency, we conducted experiments involving speech manipulations at varying levels, including sentence- and word-level deletions and substitutions. 
As shown in Fig.~\ref{fig:correlationAnalysis}, simple correlation metrics exhibit inconsistent sensitivity across forgery granularities. While the time-domain Pearson correlation drops significantly under large-scale temporal disruptions (e.g., Sentence Deletion), it fails to distinguish content substitutions, even at the sentence level (Sentence Replaced), as long as the temporal envelope remains similar. Furthermore, it remains completely stable for fine-grained word-level modifications.
This indicates that while mmWave and audio share a common physical origin, simple correlation metrics cannot reveal fine-grained tampering.
To address this, we propose a contrastive learning framework that captures more expressive cross-modal representations for robust audio authenticity verification.

\vspace{-0.05in}

\section{Threat Model}
\label{sec:threat}
We consider scenarios requiring strict voice authenticity and content integrity, such as secure communication, media forensics, and online oral examinations. 
\SystemName verifies audio authenticity and content integrity (i.e., whether the audio faithfully represents what a real human spoke at capture time) by checking cross-modal coherence between audio and mmWave-sensed throat vibrations.
Cryptographic signatures alone are insufficient, as they protect post-capture bit integrity but do not bind content to a physical source---a compromised recorder can inject forged audio and still sign it validly.
We assume an adaptive adversary utilizing forged audio to bypass detection.
They can only inject audio through digital or physical replay channels, while the radar front-end and sensing pipeline are trusted and inaccessible.
This trusted-sensor assumption is standard in capture-time integrity systems~\cite{shahid2023my,duan2024f2key}: if the acquisition hardware itself is compromised, any modality---visual, acoustic, or RF---can be bypassed. In practice, the raw mmWave data is processed on-device to produce a session-bound integrity token, and is never exposed externally.

\vspace{-0.1in}

\subsection{Attacker Capabilities}
We assume a knowledgeable adversary with knowledge of the system architecture but no access to private model weights. Specifically, the attacker possesses:

\textbf{System Knowledge:} The adversary understands the cross-modal verification mechanism (i.e., audio-throat coherence) but cannot tamper with the trained weights.

\textbf{Audio Forgery:} The adversary can utilize generative AI (e.g., TTS) or editing tools to synthesize high-fidelity, semantically consistent fake speech.

\textbf{Physical Execution:} The adversary can employ loudspeakers to simulate live speech. However, they are physically constrained: they cannot replicate non-rigid throat dynamics or actively spoof synchronized radar echoes.

\vspace{-0.05in}

\subsection{Attack Vectors}
Based on the injection pathway, we consider two vectors:

\textbf{Digital Injection:}
The adversary injects forged audio via software tools (e.g., virtual sound cards, API hijacking). While the audio channel carries forged contents, the radar observes only environmental noise. \SystemName detects this by identifying a significant modality absence.

\textbf{Physical Replay:}
The adversary plays forged audio via loudspeakers. Although loudspeakers produce convincing sound, their vibration sources (rigid diaphragms) fundamentally differ from the non-rigid vibrations of the human throat. \SystemName leverages this physical mismatch to distinguish mechanical playback from genuine human speech.

\noindent\textbf{Out-of-scope attacks.}
A copy-paste attack that replaces both the audio and mmWave streams with another matched pair compromises the capture/input channel---a general limitation of multi-sensor integrity systems.
Similarly, segment shuffling within the same recording breaks audio-mmWave time alignment; our overlapping sliding windows expose boundary mismatches across adjacent segments, making such reordering detectable.

\vspace{-0.05in}

\section{System Design}
\label{sec:design}
As shown in Fig.~\ref{fig:overview}, \SystemName\ consists of four modules: micro-vibration extraction, audio processing, spectrogram generation, and forgery detection.

\vspace{-0.1in}

\subsection{Throat Vibration Sensing}
\label{sec:mmWaveSensing}
The mmWave radar transmits linear frequency-modulated chirps and receives echoes reflected from the subject’s throat. By mixing the transmitted chirp with the conjugate of the received echo, the radar obtains an intermediate-frequency (IF) signal whose phase directly encodes the round-trip delay, and thus the instantaneous distance to the throat surface. 

To accurately localize the vibrating throat region, we perform 2D spectral analysis on the IF signal, using range FFT for distance estimation and a second FFT along the slow-time axis to capture Doppler responses from subtle movements.
Throat vibration frequencies during vocalization typically range from 100 to 200~Hz~\cite{oliveira2021fundamental}. 
We identify the throat’s approximate location in the range profile, then analyze the Doppler profile within the selected distance bin, focusing on the expected velocity-related frequency shift. The optimal range-Doppler bin matching throat vibration characteristics is selected for phase extraction.
Reflections from nearby body parts (e.g., jaw, collarbone) are suppressed because their vibration signatures fall outside the 100--200~Hz vocal range or appear in different range bins, and the dominant range-Doppler bin selection inherently isolates the strongest speech-correlated source.

Once the vibrating region is identified, we extract fine-grained vibration signals from the phase of the IF signal:
\begin{equation}
\footnotesize
\setlength\abovedisplayskip{0.1cm}
\setlength\belowdisplayskip{0.1cm}
\label{eq:phi_if}
    \phi_{\mathrm{IF}}(t)=2\pi ( f_0\tau + kt\tau - \frac{1}{2}k\tau^2 ),
\end{equation}
where $f_0$ denotes the starting frequency, and $k$ is the frequency slope of the chirp signal. Setting \(t=0\) yields the initial phase \(\phi_0\), determined by the round-trip delay \(\tau\). With \(\tau = 2d/c\), the displacement variation between adjacent chirps can be computed from their initial phases \(\phi_n\) and \(\phi_{n+1}\):
\begin{equation}
\footnotesize
\setlength\abovedisplayskip{0.15cm}
\setlength\belowdisplayskip{0.15cm}
\label{eq:delta_d}
\begin{split}
\Delta d=\frac{c}{2k}\left(\sqrt{f_0^2-\frac{k}{\pi}\phi_{n+1}}-\sqrt{f_0^2-\frac{k}{\pi}\phi_{n}}\right).
\end{split}
\end{equation}
Eq.~\ref{eq:delta_d} directly estimates inter-chirp micro-displacement, with precision determined by phase accuracy.

\begin{figure}
    \centering
    \includegraphics[width=0.95\linewidth]{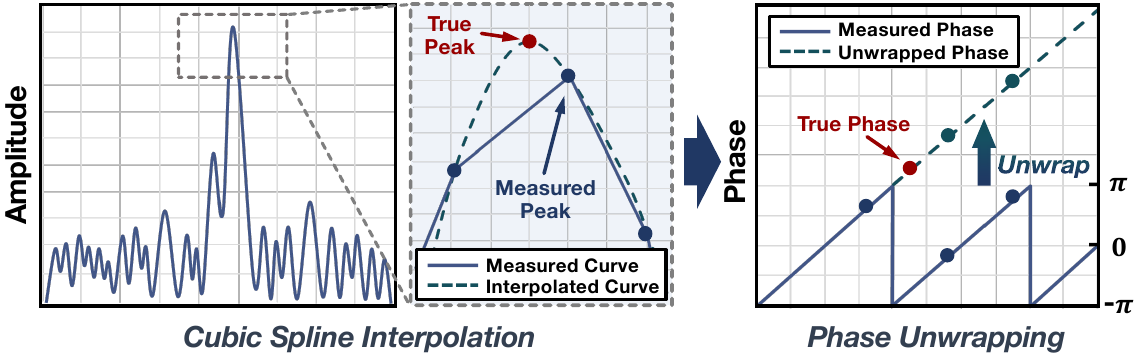}
    \vspace{-1em}
    \caption{Cubic spline interpolation and phase unwrapping.}
    \label{fig:interpolation}
\end{figure}

\begin{algorithm}[t]
\footnotesize
\caption{Composite Phase Correction}
\label{alg:phase_correction}
\begin{algorithmic}[1]
\Require Coarse phase $\phi_f$, Neighboring phases $\Phi_{\text{raw}}=\{\phi_{i-1}, \phi_i, \phi_{i+1}\}$
\Ensure Refined phase $\phi_m^{'}$

\State \textbf{Stage 1: Spectral phase refinement}
\State $\Phi_{\text{unwrap}} \leftarrow \textsc{Unwrap}(\Phi_{\text{raw}})$ \Comment{Fix discontinuities via Eq.~\ref{eq:unwrap}}
\State $\phi_m \leftarrow \textsc{CubicSplineInterp}(\Phi_{\text{unwrap}}, f_{\text{peak}})$ \Comment{Interpolate at peak}

\State \textbf{Stage 2: Residual fusion}
\State $\Delta \phi \leftarrow \phi_f - \phi_m$ \Comment{Compute residual}
\State $N_{\text{wrap}} \leftarrow \left\lfloor (\Delta \phi + \pi) / (2\pi) \right\rfloor$ \Comment{Determine wrap count}
\State $\phi_m^{'} \leftarrow \phi_m + (\Delta \phi - 2\pi \cdot N_{\text{wrap}})$ \Comment{Fuse via Eq.~\ref{eq:final_phi}}

\State \Return $\phi_m^{'}$
\end{algorithmic}
\end{algorithm}

\subsubsection{Composite Phase Correction}
Micro-displacement estimation relies on precise phase measurements. However, raw phase values \(\phi_n\) suffer from wrapping ambiguity and noise, especially under low signal-to-noise ratio (SNR) or rapid vibrations, which compromises the accuracy of Eq.~\ref{eq:delta_d}. 

To address this issue, we propose a composite phase correction algorithm that fuses coarse frequency-inferred phase with fine-grained spectral phase (see Alg.~\ref{alg:phase_correction}). 
The coarse phase \(\phi_f\) is computed from the distance estimate \(d=\frac{f_{\text{IF}}\cdot c}{2k}\), yielding round-trip delay \(\tau=\frac{2d}{c}\), substituted into Eq.~\ref{eq:phi_if} to derive \(\phi_f=2\pi n +\phi^{'}\), where \(n\) is the phase wrapping count and \(\phi^{'}\) is the fractional residual.
To refine this estimate, we leverage high-resolution spectral phases from three frequency bins centered at the IF peak, denoted as \((f^{\text{IF}}_{i-1}, f^{\text{IF}}_{i}, f^{\text{IF}}_{i+1})\) with corresponding phases \((\phi_{i-1}, \phi_i, \phi_{i+1})\). Since spectral phases are inherently wrapped within \([-\pi, \pi]\), we apply phase unwrapping to correct discontinuities between adjacent bins:
\begin{equation}
\footnotesize
\setlength\abovedisplayskip{0.15cm}
\setlength\belowdisplayskip{0.15cm}
\label{eq:unwrap}
\operatorname{unwrap}(\phi_i) = \phi_{i-1} + \left[ (\phi_i - \phi_{i-1} + \pi) \bmod 2\pi \right] - \pi.
\end{equation}
This operation ensures phase continuity between neighboring
bins.
Based on the unwrapped values, a cubic spline interpolation is then applied to estimate the refined phase $\phi_m$ at the target frequency, as illustrated in Fig.~\ref{fig:interpolation}.
Finally, we define the phase residual as \(\Delta \phi = \phi_f - \phi_m\) and apply a fusion strategy to combine the coarse phase \(\phi_f\) with the refined phase \(\phi_m\). The interpolated fine phase \(\phi_m'\) is computed as
\begin{equation}
\footnotesize
\setlength\abovedisplayskip{0.15cm}
\setlength\belowdisplayskip{0.15cm}
\label{eq:final_phi}
\phi_m^{'} = \phi_m + \left\{ \Delta \phi - 2 \pi \cdot \left \lfloor \frac{\Delta \phi + \pi}{2\pi} \right \rfloor \right\}.
\end{equation}
Substituting \(\phi_m'\) into Eq.~\ref{eq:delta_d} yields a more accurate estimation of throat micro-displacement. 

\begin{figure}
\centering
\subfigure[Time-domain signal.]{
    \begin{minipage}[t]{0.47\linewidth}
    \centering
    \includegraphics[width=1\linewidth]{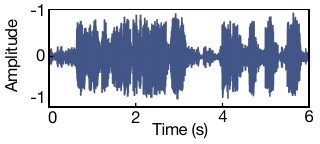}
    \vspace{-0.1in}
    \label{fig:mmw_t2}
    \end{minipage}
}
\subfigure[Time–frequency representation.]{
    \begin{minipage}[t]{0.47\linewidth}
    \centering
    \includegraphics[width=1\linewidth]{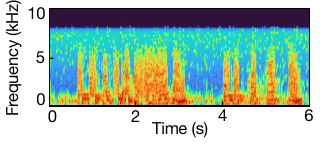}
    \vspace{-0.1in}
    \label{fig:mmw_f2}
    \end{minipage}
}
\centering
\vspace{-0.15in}
\caption{Final output of throat vibration extraction.}
\label{fig:mmw}
\end{figure}

\begin{figure*}[t!]
    \centering
    \includegraphics[width=1\linewidth]{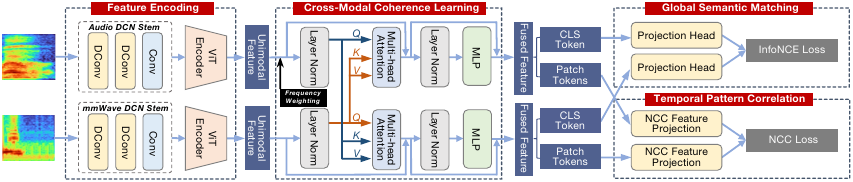}
    \vspace{-0.25in}
    \caption{The proposed CLIP-inspired \uline{C}ross-modal \uline{C}ontrastive \uline{C}oherence \uline{Net}work ($\text{C}^3$-Net).}
    \label{fig:network}
    \vspace{-0.15in}
\end{figure*}

\subsubsection{Idle-Time Interpolation for Frame-to-Frame Continuity} 
Each radar frame contains a fixed number of chirps sampled at the slow-time rate, but the radar control system inserts an idle gap between frames for RF settling and data transfer. 
Untreated, this gap disrupts the uniform sampling grid, causing spectral leakage and misalignment with the synchronous audio track in our time-domain vibration analysis.
To restore uniformity, we apply a zero-order-hold interpolation:
\begin{equation}
\footnotesize
\setlength\abovedisplayskip{0.15cm}
\setlength\belowdisplayskip{0.1cm}
    d[n]=
    \begin{cases}
    \hat d_{f,k}, & n = f(N_{\mathrm{chirp}}+1)+k, \\[2pt]
    \hat d_{f,N_{\mathrm{chirp}}}, & n = f(N_{\mathrm{chirp}}+1)+N_{\mathrm{chirp}},
    \end{cases}
\end{equation}
where $\hat d_{f,k}$ $(k=1,\dots,N_{\mathrm{chirp}})$ is the distance estimated for the $k$-th chirp in the $f$-th frame and the appended sample copies the last valid value to represent the idle gap.  
This scheme maintains phase continuity and yields a uniformly spaced signal suitable for downstream processing.
This restoration of a uniform sampling grid is critical not only for spectral purity but also for mitigating hardware clock drift.
Combined with our synchronization strategy (Sec.~\ref{sec:alig} and Sec.~\ref{sec:loss}), this ensures precise segment-level alignment without requiring complex hardware synchronization.

\subsubsection{Band-Pass Filtering and Micro-Vibration Extraction}
To eliminate respiratory drift and radar noise, the concatenated displacement series is processed using a Kaiser-windowed linear-phase FIR band-pass filter with an order of 100 and a passband of 80~Hz to 8~kHz, aligned to the device's slow-rate sampling frequency. The filtered signal is then differenced, and impulsive spikes exceeding 25~\(\mu\)m are clipped to suppress residual phase wraps and range-bin switches. Finally, the signal is peak-normalized to the range \([-1,1]\). The resulting sequence preserves sub-millimeter throat vibrations covering the fundamental and formant frequencies of speech. 
Fig.~\ref{fig:mmw} illustrates the time-domain waveforms and spectrograms of mmWave radar-extracted throat vibration signals. 

\subsection{Audio Signal Processing}
\begin{figure}[t]
\centering
\subfigure[Processed audio waveform.]{
    \begin{minipage}[t]{0.47\linewidth}
    \centering
    \includegraphics[width=0.99\linewidth]{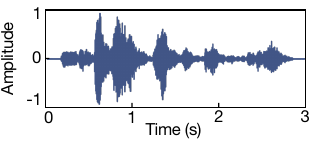}
    \vspace{-0.1in}
    \label{fig:audio_t2}
    \end{minipage}
}
\subfigure[Spectrogram of processed audio.]{
    \begin{minipage}[t]{0.47\linewidth}
    \centering
    \includegraphics[width=0.99\linewidth]{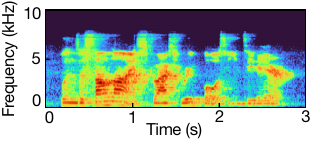}
    \vspace{-0.1in}
    \label{fig:audio_f2}
    \end{minipage}
}
\centering
\vspace{-0.15in}
\caption{Audio signal processing.}
\vspace{-0.2in}
\label{fig:audio}
\end{figure}

To enable cross-modal consistency learning, we preprocess the audio waveform to remove irrelevant components (e.g., DC offset, noise) and enhance the speech signal for time-frequency analysis. First, we remove DC bias by subtracting the mean, centering the waveform around zero for stable filtering and spectral transformations. 
Next, a Butterworth bandpass filter (80~Hz–8000~Hz) extracts speech energy, filtering out low- and high-frequency interference. 
Then, a pre-emphasis filter boosts high frequencies, sharpening formants and improving spectrogram resolution for forgery detection, applied only to audio as mmWave high-frequencies are down-weighted (Sec.~\ref{sec:F_Weighting}). 
Finally, peak normalization scales the audio signal to [-1, 1], reducing amplitude variability. 
Fig.~\ref{fig:audio} shows the processed time-domain waveforms and spectrograms of the audio signal.

\vspace{-0.1in}

\subsection{Alignment and Spectrogram Generation}
\label{sec:alig}
\subsubsection{Onset Detection}
Both mmWave and audio signals are first resampled to a unified rate to ensure consistent temporal resolution. We then compute the analytic envelope of the input signal $x(t)$ via the Hilbert transform, defined as
$ e(t) = \left| x(t) + j \cdot \mathcal{H}\{x(t)\} \right|$,
where $\mathcal{H}\{\cdot\}$ is the Hilbert transform and $j$ is the imaginary unit. This provides a phase-aware amplitude profile for speech activity detection.

To enhance the robustness of onset detection and suppress impulsive noise, we first smooth the envelope using a sliding-window median filter.
A dynamic threshold $\theta$ is then computed as $\theta = \min(e) + \alpha \cdot \left( \max(e) - \min(e) \right)$,
where $\alpha$ is a modality-specific sensitivity coefficient that adapts the threshold to different signal characteristics.
The onset is identified where the smoothed envelope exceeds $\theta$ for at least a minimum duration (e.g., 10 ms), ensuring that brief fluctuations do not trigger false positives. 

\subsubsection{Head Padding and Tail Trimming}
To prevent the loss of critical speech-initial components due to strict thresholding, we prepend a fixed temporal buffer (e.g., 20~ms) to the detected onset. 
If the buffer exceeds the signal start, it is clipped accordingly, ensuring preservation of features like glottal stops or soft consonants.
The speech offset is determined by reverse-scanning the envelope. 
When the envelope stays below a small threshold (e.g., 1.5\% of the dynamic range) for over 5~ms, silence is detected and a short buffer is added to preserve natural trailing edges.

\subsubsection{Spectrogram Generation}
After temporal alignment, both audio and mmWave signals are converted into a time– frequency representation using the Mel spectrogram. 
This representation preserves local temporal structure while emphasizing perceptually relevant frequency components, making it suitable for downstream contrastive learning framework.
Specifically, we first apply short-time Fourier transform (STFT) with a Hamming window to mitigate spectral leakage. 
The resulting spectrogram is then mapped onto the Mel scale using a triangular filter bank.
To compress the dynamic range and improve numerical stability, we apply logarithmic scaling to the Mel spectrogram with a small additive constant to avoid numerical issues. 
Both modalities are processed using identical spectral parameters, ensuring consistent representations for the cross-modal encoder.
In practice, we group consecutive Mel frames into fixed-length segments of 300\,ms and evaluate audio-mmWave coherence at this segment level rather than over entire utterances, providing word-scale sensitivity to localized tampering.

\vspace{-0.1in}

\subsection{Forgery Detection}
We introduce $\text{C}^3$-Net, a novel CLIP-inspired~\cite{radford2021learning} framework for word-level audio forgery detection, as shown in Fig.~\ref{fig:network}. 
It leverages Deformable Convolutional Network (DCN)~\cite{Zhu_2019_CVPR} stems and a Vision Transformer (ViT)~\cite{dosovitskiyimage} backbone for feature extraction. 

\vspace{-0.1in}

\subsubsection{Modality-Specific Feature Encoding} 
To capture the distinct properties of audio and mmWave signals and the local-global complexity of spectrograms, we employ a two-stage modality-specific encoder: a DCN stem for local adaptation and a ViT backbone for global context modeling.

\textbf{DCN Stems.}
The fixed receptive fields of CNNs limit their effectiveness on spectrograms with non-rigid, variable structures.
To address this, we adopt DCNs as the primary feature extractors. 
In both standard and deformable convolution, given an input feature map $\mathbf{X}$ and a fixed sampling grid $\mathcal{R}$, the output at location $\mathbf{p}_0$ is computed as
\begin{equation}
\footnotesize
\setlength\abovedisplayskip{0.15cm}
\setlength\belowdisplayskip{0.1cm}
\mathbf{Y}(\mathbf{p}_0) = \sum_{\mathbf{p}_k \in \mathcal{R}} \mathbf{w}(\mathbf{p}_k) \cdot \mathbf{X}(\mathbf{p}_0 + \mathbf{p}_k + \Delta \mathbf{p}_k),
\end{equation}
where $\mathbf{w}(\mathbf{p}_k)$ denotes the kernel weight at position $\mathbf{p}_k$, and $\Delta \mathbf{p}_k$ is a 2D offset. When $\Delta \mathbf{p}_k = \mathbf{0}$, this reduces to the standard convolution. 
By enabling adaptive sampling, DCNs adjust receptive fields on demand and better capture complex harmonics and time-varying formants.

\begin{figure}[t!]
    \centering
    \includegraphics[width=0.99\linewidth]{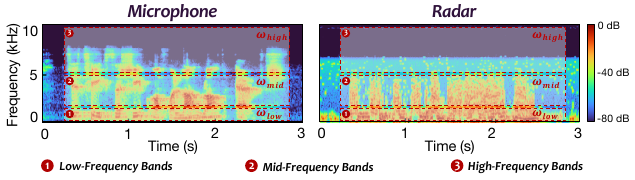}
    \vspace{-0.15in}
    \caption{Cross-modal frequency band energy comparison.}
    \label{fig:fWeighting}
    \vspace{-0.15in}
\end{figure}

To accommodate the distinct characteristics of the two modalities (see Fig.~\ref{fig:fWeighting}), we adopt an asymmetric DCN design.
For audio spectrograms with rich textures and high-frequency details, a two-layer $3\times3$ DCN is used: the first layer captures basic time-frequency patterns, while the second, with more deformable groups, traces complex harmonics and formants.
In contrast, mmWave signals, dominated by coarse low- to mid-frequency patterns, are processed with a two-layer $5\times5$ DCN, where the larger kernel improves noise robustness and captures broader energy regions.

\textbf{ViT Encoders.}
Afterwards, the feature maps are fed to a ViT encoder for deep contextual learning. 
The ViT splits each feature map $\mathbf{F}$ into $N$ fixed-size patches ${\mathbf{f}_p^i}$. These patches are flattened and projected into embedding tokens via a trainable matrix $\mathbf{E}$. A learnable class token $\mathbf{z}_{\text{cls}}$ and position embeddings $\mathbf{E}_{\text{pos}}$ are added, forming the input sequence:
\begin{equation}
\footnotesize
\setlength\abovedisplayskip{0.15cm}
\setlength\belowdisplayskip{0.15cm}
\mathbf{z}_0 = [\mathbf{z}_{\text{cls}};  \mathbf{f}_p^1 \mathbf{E};  \mathbf{f}_p^2 \mathbf{E};  \dots;  \mathbf{f}_p^N \mathbf{E}] + \mathbf{E}_{\text{pos}}.
\end{equation}
This sequence passes through $L$ Transformer blocks, each with Multi-head Self-Attention (MSA) and Multi-Layer Perceptron (MLP) modules, linked by residual connections and Layer Normalization. ViT’s self-attention captures global dependencies, essential for relating distant spectro-temporal features in spectrograms. 
The output encodes both local and global information, with the class token embedding $\mathbf{z}_L^0$ representing a condensed global modality-aware feature. 
This output feeds into the CMA module for deep fusion.

\subsubsection{Frequency-Aware Patch Weighting}
\label{sec:F_Weighting}
To further incorporate the physical prior knowledge of mmWave signals and enhance model robustness, we propose a Frequency-Aware Patch Weighting mechanism. This design is motivated by empirical observations: as shown in Fig.~\ref{fig:fWeighting}, in the mmWave spectrogram, effective physical vibration signals corresponding to actual speech are predominantly concentrated in the low- and mid-frequency bands, whereas the high-frequency regions tend to be sparse and more susceptible to noise interference. Treating all frequency bands equally may therefore degrade the quality of the learned representations.

We apply frequency-dependent weights to the patch tokens of the mmWave ViT output $\mathbf{S}_{\text{mmw}} = [\mathbf{z}_L^0, \mathbf{z}_L^1, \dots, \mathbf{z}_L^N]$ before cross-modal fusion. 
Each patch token $\mathbf{z}_L^i$ is reweighted as $w_i \in \{{\omega_{\text{low}}, \omega_{\text{mid}}, \omega_{\text{high}}}\}$ based on its row index $r(i)$ compared to thresholds $R_{\text{low}}$ and $R_{\text{mid}}$.
Here, $R_{\text{low}}$ and $R_{\text{mid}}$ define the frequency band boundaries, and $(\omega_{\text{low}}, \omega_{\text{mid}}, \omega_{\text{high}})$ denote fixed weights set empirically to (0.7, 1.0, 0.3) based on the energy distribution analysis in Fig.~\ref{fig:fWeighting}.
The new weighted feature sequence is then constructed by applying these weights to the patch tokens while leaving the CLS token unchanged to preserve global semantics:
\begin{equation}
\footnotesize
\setlength\abovedisplayskip{0.15cm}
\setlength\belowdisplayskip{0.15cm}
    \hat{\mathbf{S}}_{\text{mmw}} = [\mathbf{z}_L^0, w_1 \cdot \mathbf{z}_L^1, w_2 \cdot \mathbf{z}_L^2, \dots, w_N \cdot \mathbf{z}_L^N].
\end{equation}
This weighting guides the model to attend more to informative, noise-robust frequency regions during CMA.

\subsubsection{Cross-Modal Coherence Learning}
Conventional fusion methods, such as concatenation or summing final embeddings, fail to capture fine-grained, token-level cross-modal correspondences. 
To overcome this, we propose a CMA module enabling deep, bidirectional interaction between audio and frequency-weighted mmWave features. It comprises two parallel attention branches with opposite query directions:

\begin{figure}[t!]
    \centering
    \includegraphics[width=1\linewidth]{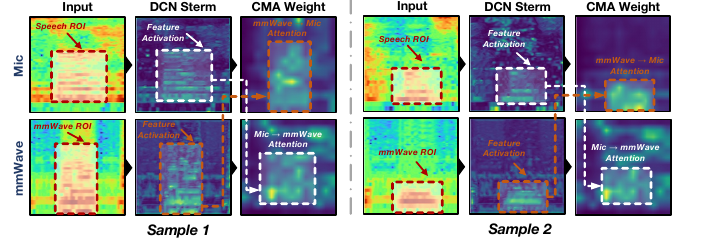}
    \vspace{-20pt}
    \caption{Intermediate feature maps of the cross-modal attention.}
    \label{fig:CMAFeatures}
    \vspace{-0.2in}
\end{figure}

\textit{1) Audio-to-mmWave Attention:} Audio tokens serve as queries, attending to mmWave tokens as keys and values. This allows each audio token to integrate relevant physical cues from the mmWave stream.

\textit{2) mmWave-to-Audio Attention:} mmWave tokens act as queries and attend to the audio tokens, refining their representations with extra spectral and harmonic details.

Through fine-grained token-level interactions, the CMA module learns localized, content-aware alignments between different modalities, as shown in Fig.~\ref{fig:CMAFeatures}. For instance, audio harmonic structures often correspond to specific vibration energy clusters in mmWave data.

\subsubsection{Training Objective and Loss Function Design} 
\label{sec:loss}
We optimize $\text{C}^3$-Net with two objectives: global semantic alignment via contrastive learning and local temporal coherence.

\textbf{Global Contrastive Loss.}
The primary objective is to build a shared embedding space where audio and mmWave signals originating from the same real-world event are treated as a positive pair and are pulled together, while samples from different events serve as negative samples and are pushed apart.
We use a symmetric InfoNCE loss~\cite{oord2018representation} to align audio and mmWave embeddings.
Let $a_i$ and $m_i$ be the L2-normalized embeddings of the $i$-th audio and mmWave samples, respectively. The loss is defined as
\begin{equation}
\footnotesize
\setlength\abovedisplayskip{0.15cm}
\setlength\belowdisplayskip{0.15cm}
\mathcal{L}_{\text{I}} = -\frac{1}{2N} \sum_{i=1}^{N} \left[
    \log \frac{e^{a_i \cdot m_i / \kappa}}{\sum_{j=1}^{N} e^{a_i \cdot m_j / \kappa}} 
    + \log \frac{e^{m_i \cdot a_i / \kappa}}{\sum_{j=1}^{N} e^{m_i \cdot a_j / \kappa}}
\right],
\end{equation}
where $a_i \cdot m_j$ is the cosine similarity between the $i$-th audio and $j$-th mmWave embeddings, and $\kappa$ is a learnable temperature parameter controlling the sharpness of the similarity distribution.
This encourages the model to ignore low-level modality-specific idiosyncrasies, and focus on high-level semantic fingerprints that signify the same real-world event.

\textbf{Temporal Coherence Loss.}
To complement the global semantic alignment, we introduce an auxiliary loss that enforces local temporal coherence between audio and mmWave modalities. This loss ensures the model captures fine-grained temporal patterns, improving robustness to subtle misalignments that may indicate forgery.
Specifically, we maximize the Normalized Cross-Correlation (NCC) between 1D temporal feature sequences from both modalities. To handle slight timing differences, we search over a small lag window and take the maximum NCC value as the coherence measure.
Let $\mathbf{t}_{\text{mic}}$ and $\mathbf{t}_{\text{mmw}}$ denote the temporal sequences from audio and mmWave, respectively. The loss is defined as
\begin{equation}\footnotesize
\setlength\abovedisplayskip{0.1cm}
\setlength\belowdisplayskip{0.13cm}
\mathcal{L}_{\text{N}} = 1 - \mathbb{E} \left[ \max_{k \in [-\lambda_{\text{lag}}, \lambda_{\text{lag}}]} \text{NCC}(\mathbf{t}_{\text{mic}}, \mathbf{t}_{\text{mmw}}, k) \right],
\end{equation}
where $\text{NCC}(\cdot, \cdot, k)$ is the normalized cross-correlation at lag $k$, and $\lambda_{\text{lag}}$ defines the lag search range. 
This objective encourages the model to learn feature representations whose temporal evolution patterns are tightly synchronized, thereby capturing a crucial physical property of genuine speech.

\textbf{Total Loss.}
We combine the primary and auxiliary objectives in a multi-task learning framework. The overall loss is given by 
$\mathcal{L} = \mathcal{L}_{\text{I}} + \lambda \cdot \mathcal{L}_{\text{N}}$,
where $\lambda$ is a weighting hyperparameter controlling temporal coherence.

\vspace{-0.15in}
\section{Evaluation}
\label{sec:evaluation}
\subsection{Experimental Setup}
\subsubsection{Devices}
\begin{figure}
    \centering
    \includegraphics[width=0.9\linewidth]{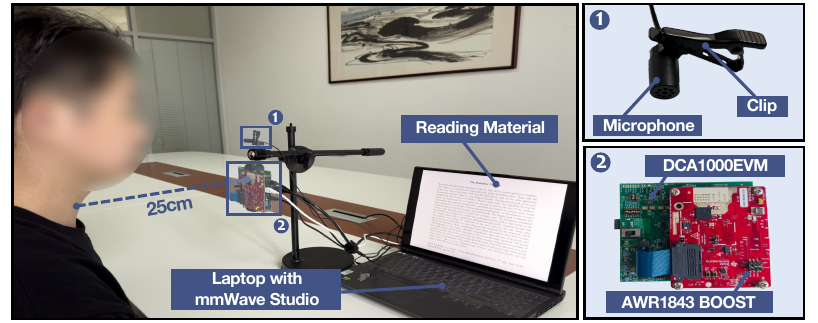}
    \vspace{-0.1in}
    \caption{Experimental setup.}
    \label{fig:expset}
\end{figure}

Our system employs a TI AWR1843 mmWave radar~\cite{TI_AWR1843} and a BY-M1S clip-on microphone~\cite{BOYA_BY‑M1S}, both mounted on the same support structure and positioned 25~cm in front of the speaker, as illustrated in Fig.~\ref{fig:expset}. The radar operates at a center frequency of 77~GHz with 4~GHz bandwidth and a chirp duration of 60~$\mu$s, yielding a range resolution of 3.75~cm. The frame rate is set to 1000~fps to capture fast vocal dynamics. A laptop runs mmWave Studio and Matlab 2024b for data collection. 
Our model is implemented in Python~3.9 using PyTorch~2.7 and trained on a PC equipped with an NVIDIA GeForce RTX 4090 GPU.

\begin{figure*}[t]
\centering
\subfigure[ROC curve.]{
    \begin{minipage}[t]{0.235\linewidth}
    \centering
    \includegraphics[width=1\linewidth]{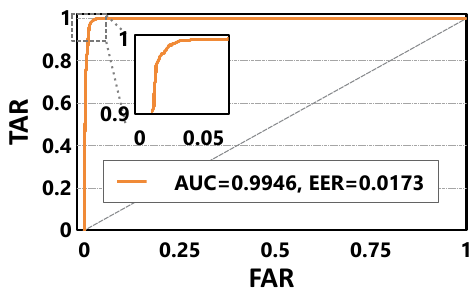}
    \vspace{-14pt}
    \label{fig:roc}
    \end{minipage}
}
\subfigure[Score distribution.]{
    \begin{minipage}[t]{0.235\linewidth}
    \centering
    \includegraphics[width=1\linewidth]{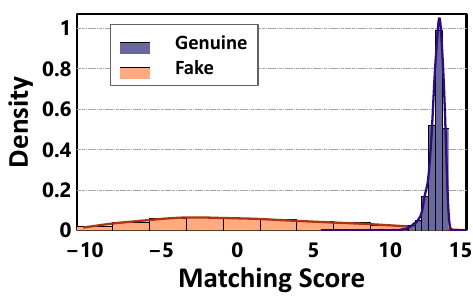}
    \vspace{-14pt}
    \label{fig:score}
    \end{minipage}
}
\subfigure[TAR.]{
    \begin{minipage}[t]{0.235\linewidth}
    \centering
    \includegraphics[width=1\linewidth]{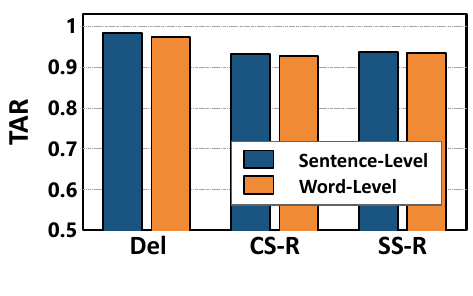}
    \vspace{-14pt}
    \label{fig:attackTAR}
    \end{minipage}
}
\subfigure[FAR.]{
    \begin{minipage}[t]{0.235\linewidth}
    \centering
    \includegraphics[width=1\linewidth]{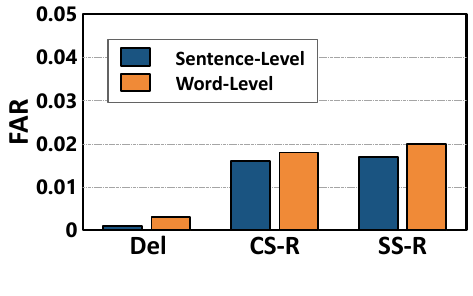}
    \vspace{-14pt}
    \label{fig:attackFAR}
    \end{minipage}
}
\centering
\vspace{-0.2in}
\caption{Overall performance, including global discrimination metrics and robustness against specific forgery types.}
\label{fig:overallExperiment}
\vspace{-0.15in}
\end{figure*}

\begin{table}[t]
\centering
\caption{Summary of reading passages.}
\vspace{-0.10in}
\label{tab:readpassages}
\resizebox{\columnwidth}{!}{
\begin{tabular}{lccc}
\toprule
\textbf{Passage Title} & \makecell{\textbf{English} \\ \textbf{Word Count}} & \makecell{\textbf{Chinese} \\ \textbf{Word Count}} & \textbf{Content Type} \\
\midrule
The Rainbow Passage & 329 & 497 & Scientific Description \\
The North Wind and the Sun & 209 & 303 & Fable \\
The Grandfather Passage & 132 & 226 & Character Narrative \\
Arthur the Rat & 335 & 559 & Animal Story \\
\bottomrule
\end{tabular}
}
\vspace{-0.25in}
\end{table}

\subsubsection{Datasets}
We collected synchronized audio and radar data from 21 volunteers (12 male, 9 female)\footnote{Ethical approval has been obtained from the corresponding organization.} reading a fixed set of texts, which cover a variety of phonetic and linguistic structures, commonly used in speech processing benchmarks (see Tab.~\ref{tab:readpassages}).
To ensure physical robustness, participants were selected with diverse body types, capturing natural variability in throat geometry.
This original dataset is used to train the model’s cross-modal alignment capability.
For the primary evaluation, we adopt a random utterance-level split (80\% training, 20\% testing), ensuring that the model learns baseline physical coherence from a broad set of genuine speech samples.
In addition, to assess the model's ability to generalize to unseen users, we further perform a speaker-independent split: we randomly select $N_\text{test}$ volunteers (e.g., 4) as the test set and use the remaining volunteers for training.
All utterances from the test speakers are strictly held out during training.

To evaluate performance under realistic speech tampering scenarios, we construct two representative adversarial testing datasets, based on the held-out 20\% test utterances, using semantic transcription and temporal boundary extraction generated by the Whisper model~\cite{radford2023robust}.

\textit{1) Sentence-level Tampering Dataset:} Each utterance is segmented into sentences with corresponding start and end timestamps. A target sentence is either deleted or replaced with another complete sentence of similar duration (from the same or a different speaker).

\textit{2) Word-level Tampering Dataset:} Using word-level timestamp annotations, specific words are deleted or substituted with duration-matched alternatives from the corpus. 

The resulting forged audio is realigned with the mmWave signal and translated to Mel spectrogram windows for evaluation.
For each tampering subtype (i.e., deletion, cross-speaker replacement, same-speaker replacement), we sample 300 non-overlapping instances from the test set to ensure balanced evaluation across conditions, 
resulting in 1,800 tampered samples in total across sentence- and word-level datasets.

\vspace{-1em}
\subsubsection{Metrics}
We evaluate \SystemName's performance using three standard metrics widely used in speech tampering detection: True Acceptance Rate (TAR), False Acceptance Rate (FAR), and Equal Error Rate (EER). 
TAR and FAR reflect the system’s ability to correctly accept genuine inputs and reject tampered ones, while EER provides an overall measure of detection balance.

\vspace{-1em}

\begin{figure}[t]
\centering
\subfigure[Setup of physical replay attack.]{
    \begin{minipage}[t]{0.455\linewidth}
    \centering
    \includegraphics[width=1\linewidth]{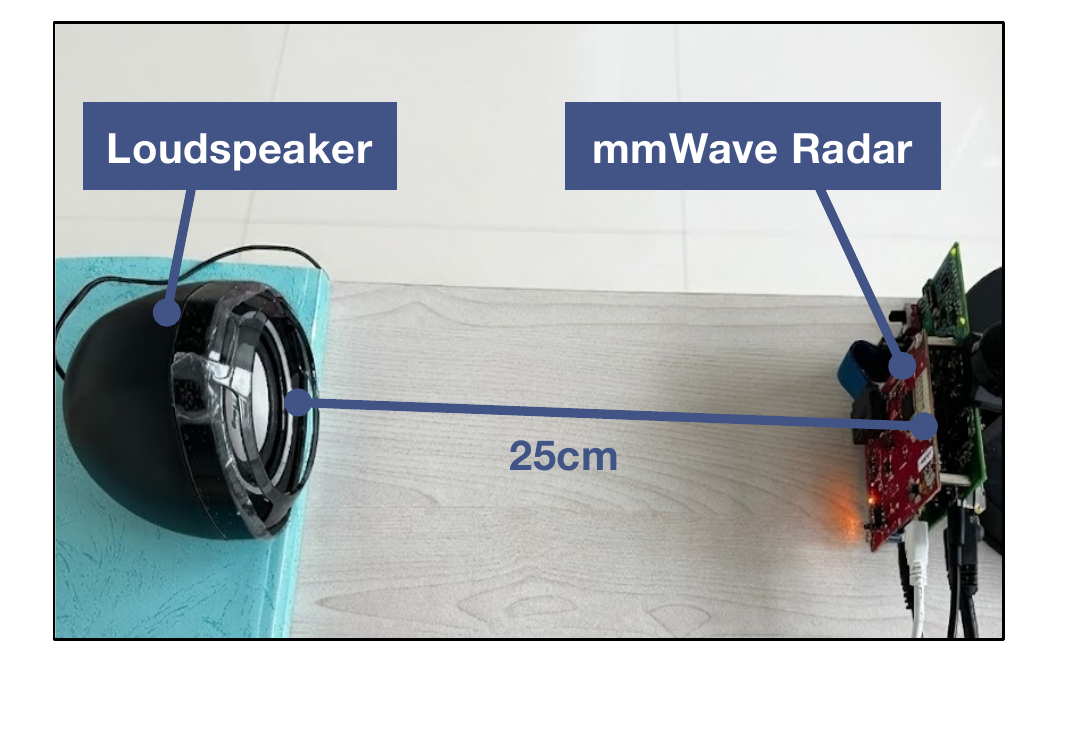}
    \vspace{-8pt}
    \label{fig:phyAttackScen}
    \end{minipage}
}\vspace{-0.7em}\hspace{-0.3em}
\subfigure[Acceptance rate.]{
    \begin{minipage}[t]{0.47\linewidth}
    \centering
    \includegraphics[width=1\linewidth]{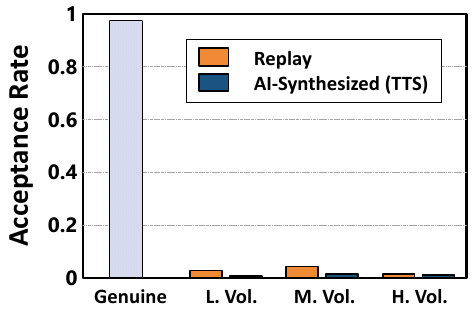}
    \vspace{-8pt}
    \label{fig:eva_phy}
    \end{minipage}
}
\subfigure[Spectral characteristics comparison.]{
    \begin{minipage}[t]{1\linewidth}
    \centering
    \includegraphics[width=.94\linewidth]{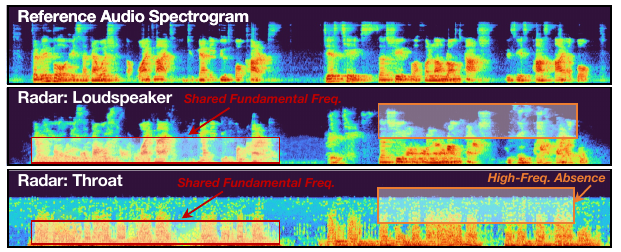}
    \vspace{-8pt}
    \label{fig:phySpe}
    \end{minipage}
}
\centering
\vspace{-0.2in}
\caption{Performance on physical replay and deepfake attacks.}
\label{fig:PhyattackExperiment}

\end{figure}

\subsection{Overall Performance}
We first evaluate the system's baseline capability in distinguishing matched genuine pairs from randomly mismatched pairs.
The coherence score is computed as the cosine similarity between the L2-normalized audio and mmWave embeddings produced by the CMA module.
As shown in Fig.\ref{fig:roc}, the system achieves strong discrimination performance, with an AUC of 0.9946 and an EER of 0.0173.
This is consistent with the score distribution in Fig.\ref{fig:score}, where genuine pairs concentrate at high coherence scores, while fake pairs spread across lower values.
The clear separation between the two distributions indicates that \SystemName can effectively capture cross-modal consistency for reliable forgery detection.

To further analyze the system's robustness against specific forgery types, we report the TAR and FAR under sentence- and word-level attacks, as shown in Fig.~\ref{fig:attackTAR} and Fig.~\ref{fig:attackFAR}.

\textit{1) Sentence-level Attacks} simulate the replacement or deletion of an entire sentence. We aggregate window-level scores using majority-voting to determine whether a sentence has been tampered with. Our system achieves a high TAR of 98.45\% for sentence deletion (Del), and 93.3\% and 93.7\% for cross-speaker (CS-R) and same-speaker (SS-R) replacement, respectively. 
The corresponding FARs remain low (0–1.7\%), reflecting the clear cross-modal inconsistency induced by sentence deletion.

\textit{2) Word-level Attacks} focus on localized manipulations such as replacing or deleting specific keywords. Our system achieves TARs of 97.42\%, 92.8\%, and 93.5\% under deletion, cross-speaker replacement, and same-speaker replacement, respectively. 
FARs remain low (<2.0\%), indicating reliable detection performance even under subtle tampering.

These experiments focus on physically embedded forgeries where both audio and radar contain strong signals but originate from mismatched physical sources. 
In contrast, pure digital-injection attacks, which lack any corresponding throat vibration, reduce to a simple modality-absence case (Sec.~\ref{sec:threat}) and are thus not further evaluated.

\begin{figure*}
\centering
\subfigure[Unseen speakers.]{
    \begin{minipage}[t]{0.23\linewidth}
    \centering
    \includegraphics[width=1\linewidth]{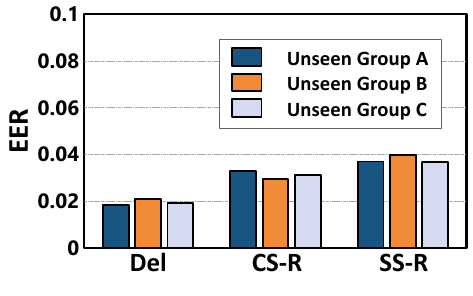}
    \vspace{-1em}
    \label{fig:speakEva}
    \end{minipage}
}
\subfigure[Different languages.]{
    \begin{minipage}[t]{0.23\linewidth}
    \centering
    \includegraphics[width=1\linewidth]{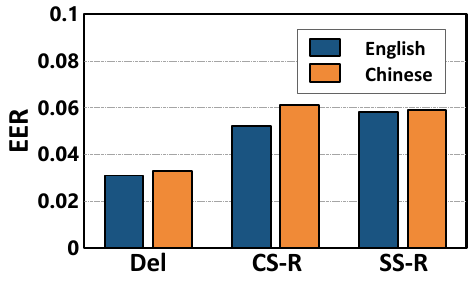}
    \vspace{-1em}
    \label{fig:langEva}
    \end{minipage}
}
\subfigure[Different distances.]{
    \begin{minipage}[t]{0.23\linewidth}
    \centering
    \includegraphics[width=1\linewidth]{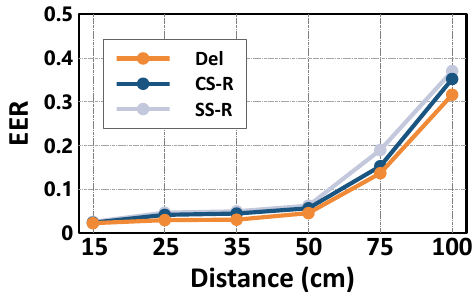}
    \vspace{-1em}
    \label{fig:disEva}
    \end{minipage}
}
\subfigure[Different angles.]{
    \begin{minipage}[t]{0.23\linewidth}
    \centering
    \includegraphics[width=1\linewidth]{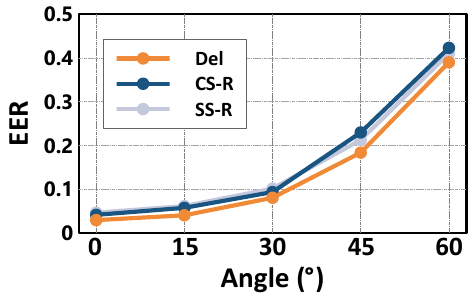}
    \vspace{-1em}
    \label{fig:angEva}
    \end{minipage}
}
\centering
\vspace{-1em}
\caption{Robustness and generalization experiments.}
\label{fig:robustExperiment}
\vspace{-1em}

\end{figure*}

\subsection{Performance on Replay and Generative Deepfakes}

We evaluated \SystemName against physically delivered attacks, including \textit{recorded speech replay} and \textit{AI-synthesized deepfakes} (generated via OpenAI TTS~\cite{ttsopenAI}).
Using a loudspeaker at 25~cm (Fig.~\ref{fig:phyAttackScen}), we simulated attacks at three volume levels.
The results are presented in Fig.~\ref{fig:eva_phy}. 
Our system maintains a high TAR ($>97\%$) for genuine users. 
For attacks, the FAR for recorded replay drops below 5\% across all volumes, while TTS attacks achieve near-zero acceptance.

To understand the physical basis of this rejection, we visualize the spectral characteristics in Fig.~\ref{fig:phySpe}.
As observed, the radar response from the loudspeaker (Middle) exhibits distinct high-frequency artifacts, effectively acting as a rigid spectral copy of the reference audio (Top).
In contrast, the human throat response (Bottom) shows significant High-Frequency Absence due to physiological damping, preserving only the shared fundamental frequency associated with the glottal source.
This contrast validates the speech production mechanism discussed in Sec.~\ref{sec:preliminary_speech}: a loudspeaker diaphragm vibrates to reproduce the entire acoustic signal, whereas the radar captures only the throat source excitation.
\SystemName detects this fundamental mismatch, where radar captures throat vibrations rather than the full acoustic signal, and reliably rejects the attack.

\begin{table}[t]
\centering
\caption{EER comparison with state-of-the-art methods.}
\vspace{-0.1in}
\label{tab:baselineComparison}
\resizebox{\columnwidth}{!}{
\begin{tabular}{lcccccc}
\toprule
\textbf{Method} & \multicolumn{3}{c}{\textbf{Sentence-level Forgery}} & \multicolumn{3}{c}{\textbf{Word-level Forgery}} \\
\cmidrule(lr){2-4} \cmidrule(lr){5-7}
& \textbf{Del} & \textbf{CS-R} & \textbf{SS-R} & \textbf{Del} & \textbf{CS-R} & \textbf{SS-R} \\
\midrule
RawNet2~\cite{tak2021end} & 0.109 & 0.183 & 0.304 & 0.173 & 0.238 & 0.345 \\
Wav2Vec2 + MLP~\cite{baevski2020wav2vec} & 0.091 & 0.159 & 0.283 & 0.143 & 0.210 & 0.356 \\
W2V-BERT2~\cite{barrault2023seamless} & 0.079 & 0.136 & 0.265 & 0.131 & 0.188 & 0.327\\
A-V Mismatch~\cite{bohacek2024lost} & 0.052 & 0.107 & 0.153 & 0.109 & 0.121 & 0.187\\
\textbf{Ours} & \textbf{0.027} & \textbf{0.033} & \textbf{0.041} & \textbf{0.030} & \textbf{0.042} & \textbf{0.047} \\
\bottomrule
\end{tabular}
}
\vspace{-0.2in}
\end{table}

\subsection{Performance Comparison}
We compare our system against representative audio-only baselines and a cross-modal baseline, including RawNet2~\cite{tak2021end}, Wav2Vec2~\cite{baevski2020wav2vec}, W2V-BERT2~\cite{barrault2023seamless}, and Audio-Video (A-V) Mismatch~\cite{bohacek2024lost}.
RawNet2 is a waveform-based CNN model designed for spoof detection. We retrain it for binary classification using our tampering data. 
Wav2Vec2 employs a self-supervised encoder pretrained on large-scale speech data. Although originally built for speech recognition, it can be adapted to forgery detection by replacing the decoder with a MLP classifier.
We additionally include W2V-BERT2 as a stronger state-of-the-art audio-only baseline by attaching a classification head to its pretrained speech representations for forgery detection. For cross-modal comparison, we include A-V Mismatch, which detects forgery by separately transcribing audio and visual lip movements and measuring their inconsistency.

As shown in Tab.~\ref{tab:baselineComparison}, the compared baselines exhibit clear limitations across different tampering types. RawNet2, operating purely in the waveform domain, is sensitive to coarse discontinuities such as sentence deletion, but degrades sharply under subtler manipulations like same-speaker replacement. Wav2Vec2 shows modest improvement due to its self-supervised phonetic representations, yet still struggles with fine-grained edits, reaching an EER of 0.356 for same-speaker replacement. Although W2V-BERT2 consistently outperforms Wav2Vec2, its performance remains limited, especially under fine-grained tampering. These results reveal a fundamental limitation of audio-only detectors: relying solely on acoustic cues, they cannot reliably expose perceptually natural but semantically inconsistent edits, nor manipulations that preserve global speech statistics. A-V Mismatch further improves over audio-only baselines by incorporating visual speech cues, but its performance remains below \SystemName. This is because lip-based verification captures only external articulatory patterns, which provide less direct evidence for fine-grained content integrity than source-level throat vibrations. Subtle manipulations, especially word-level edits and same-speaker replacements, can still preserve visually plausible lip motion, while phonation-level differences remain difficult to infer from mouth appearance alone. By contrast, \SystemName uses throat vibrations as a source-coupled physical anchor to capture inconsistencies rooted in the speech production process, enabling more reliable detection of subtle forgeries that are difficult to reveal from either audio or lip motion alone.

\subsection{Robustness and Generalization}
We evaluate the system's robustness under the word-level forgery detection task in this section.
\vspace{-0.02in}

\subsubsection{Generalization to Unseen Speakers}
We assess robustness under a speaker-independent setting by randomly forming three disjoint test groups, each containing 4 held-out speakers, with the remaining speakers used for training. This setup evaluates the model’s ability to generalize to users not seen during training.
As shown in Fig.~\ref{fig:speakEva}, the system maintains stable performance across all unseen speaker groups, demonstrating strong cross-speaker generalization. 
Specifically, word deletion (Del) yields the lowest EER (approximately 0.02), while replacement attacks (CS-R and SS-R) lead to slightly higher EERs, with SS-R being the most challenging due to its subtle intra-speaker nature.

\subsubsection{Impact of Language}
We evaluate the impact of language on forgery detection, focusing on English and Chinese. The model is trained on single-language data and evaluated on the other. As shown in Fig.~\ref{fig:langEva}, the results indicate a slight increase in overall EER due to reduced training data, with higher EER observed for Chinese. 
Specifically, the EERs for Word-Del, Word-CS-R, and Word-SS-R are 0.031, 0.052, and 0.058 in English, and 0.033, 0.061, and 0.059 in Chinese, respectively. 
Despite these differences, the system maintains good cross-language detection performance.
\vspace{-0.02in}

\subsubsection{Impact of Distance}
We evaluate the effect of speaker-to-radar distance in Fig.~\ref{fig:disEva}.
The system maintains a low EER ($<0.05$) within 50~cm, showing high sensitivity to subtle throat vibrations at close range.
Performance drops at longer distances (EER $>0.3$ at 100~cm) due to signal attenuation and beam dispersion.
This effective range matches our target scenarios, where users are naturally close to the device.
It also improves privacy by limiting analysis to nearby speakers and reducing interference from distant background voices.

\begin{figure}[t]
    \centering
    \begin{minipage}[t]{0.47\linewidth}
        \centering
        \includegraphics[width=1\linewidth]{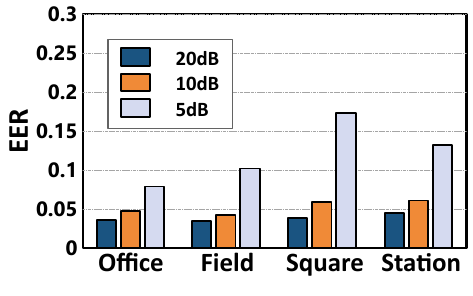}
        \vspace{-2.5em}
        \caption{Effect of noise.}
        \label{fig:noiseEva}
    \end{minipage}
    \begin{minipage}[t]{0.47\linewidth}
        \centering
        \includegraphics[width=1\linewidth]{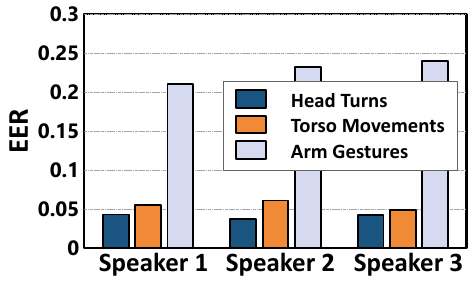}
        \vspace{-2.5em}
        \caption{Effect of motion.}
        \label{fig:motionEva}
    \end{minipage}
    \vspace{-1em}
\end{figure}

\begin{table}[t]
\centering
\caption{Ablation study on architectural elements.}
\vspace{-1em}
\label{tab:ablation}
\resizebox{\columnwidth}{!}{
\begin{tabular}{lccc}
\toprule
\textbf{Setting}  & \textbf{EER} $\downarrow$ & \textbf{Mic$\rightarrow$mmWave Accuracy} $\uparrow$ & \textbf{mmWave$\rightarrow$Mic Accuracy} $\uparrow$ \\
\midrule
w/o DCN Module       & 0.064          & \timebar{100}{95.15}\%           & \timebar{100}{96.47}\%           \\
w/o CMA Module       & 0.187           & \timebar{100}{54.30}\%          & \timebar{100}{53.78}\%          \\
w/o NCC Loss         & 0.038           & \timebar{100}{96.56}\%           &\timebar{100}{96.70}\%          \\
w/o InfoNCE Loss     & 0.484           & \timebar{100}{6.97}\%           & \timebar{100}{6.73}\%           \\
\midrule
\textbf{Full Model}           & 0.017  & \timebar{100}{99.84}\%  & \timebar{100}{99.83}\%  \\
\bottomrule
\end{tabular}}
\vspace{-0.8em}
\end{table}

\subsubsection{Impact of Angle}
We evaluate how speaker–radar angular offset affects detection performance by varying the angle from $0^\circ$ to $60^\circ$ at a fixed distance of 25~cm. 
As shown in Fig.~\ref{fig:angEva}, the EER remains low ($\leq$0.1) when the angle is within $30^\circ$, but increases significantly at larger offsets. Specifically, EER rises to 0.209 at $45^\circ$ and reaches 0.408 at $60^\circ$ due to beam dispersion and reduced signal strength. 
These results indicate near-frontal alignment is key to capturing stable throat vibrations and ensuring accurate detection.

\subsubsection{Impact of Environmental Noise}
We evaluate forgery detection under environmental noise by adding four DEMAND noises~\cite{thiemann_joachim_2013_1227121}, Office, Field, Square, and Station, to the audio signals at 20, 10, and 5~dB SNR.
As shown in Fig.~\ref{fig:noiseEva}, performance degrades as noise increases, with the average EER across the three forgery types rising to 0.173 at 5~dB.
Square and Station cause the largest degradation, likely due to their more complex acoustic characteristics.
Nevertheless, the system maintains relatively low EERs under moderate noise, indicating strong robustness.

\subsubsection{Impact of Body Motion}
We evaluate robustness under three motion conditions: slight head turns, moderate torso movements, and vigorous arm gestures.
Fig.~\ref{fig:motionEva} reports the average EER across three forgery types for each condition.
Performance remains stable under slight head motion (EER $\approx 0.041$), degrades moderately with torso movement (EER $\approx 0.055$), and drops sharply under vigorous arm gestures (EER rising to 0.228).
The decline mainly stems from arm-induced occlusion, which disrupts throat tracking and introduces signal artifacts.
Overall, the system handles mild motion well but degrades under intense movement.

\subsection{Architectural Element Study}
To evaluate the contributions of individual architectural components, we conduct an ablation study on four key elements: the DCN module, the CMA module, and the dual-loss objective consisting of InfoNCE and NCC losses.

As presented in Tab.~\ref{tab:ablation}, the full model achieves the best performance with an EER of 0.017 and a bidirectional retrieval accuracy of 99.8\%, demonstrating excellent cross-modal consistency and tampering detection capability.
Removing the DCN module slightly increases the EER to 0.064, indicating its effectiveness in capturing spatial dynamics. 
In contrast, discarding the CMA module causes a drastic performance drop, with EER rising to 0.187 and retrieval accuracy dropping to around 55\%, highlighting the importance of cross-modal interaction.
On the loss side, removing the NCC loss causes only a marginal degradation, while eliminating the InfoNCE loss severely impairs performance, resulting in an EER of 0.484 and accuracy below 7\%. This suggests that InfoNCE plays a dominant role in maintaining discriminative representations, while NCC complements it by enforcing temporal alignment.
These results confirm that both architecture and loss design are essential for system robustness.

\vspace{-0.1in}

\subsection{Real-time Capability and Overhead} 
We evaluate the real-time feasibility and computational efficiency of \SystemName. Experiments were conducted with a batch size of 1 to simulate a streaming inference scenario, and the average metrics were calculated over 1000 runs.

\begin{table}[t]
\centering
\caption{Average inference time and computation per module.}
\vspace{-1em}
\label{tab:timeOverhead}
\scriptsize
\resizebox{\columnwidth}{!}{
\begin{tabular}{lcccc}
\toprule
\textbf{Module} & \textbf{Time (ms)} & \textbf{Time Proportion} & \textbf{GFLOPs} & \textbf{GFLOPs Proportion} \\
\midrule
Audio Feature Encoding          & 4.8077  &  \timebar{100}{46.20}\% & 17.9507 &  \timebar{100}{46.64}\% \\
mmWave Feature Encoding         & 4.8032  &  \timebar{100}{46.16}\% & 19.0811 &  \timebar{100}{49.58}\%\\
CMA Module                      & 0.6857  &  \timebar{100}{6.59}\% & 1.4562 &  \timebar{100}{3.78}\% \\
Decision Stage                  & 0.1096  &  \timebar{100}{1.05}\%  & 0.0005 &       \timebar{100}{0.00}\% \\
\midrule
\textbf{Total}                  & 10.4062 & \timebar{100}{100}\%    & 38.4886 &  \timebar{100}{100}\% \\
\bottomrule
\end{tabular}}
\vspace{-2em}
\end{table}

As shown in Tab.~\ref{tab:timeOverhead}, the total inference latency for a single sample is approximately 10.41~ms. 
This latency is small compared with the duration of a typical input window for word-level detection (hundreds of milliseconds to seconds), resulting in a Real-Time Factor (RTF) well below 1.
This implies that the system can process incoming signal streams much faster than they are produced, leaving ample margin for system buffering and other concurrent tasks.
In terms of computational distribution, most of the runtime originates from audio and mmWave feature encoders, which together account for over 92\% of the total latency. 
The CMA fusion module introduces a modest overhead, and the decision stage is negligible. 
Overall, the system achieves an effective throughput of roughly 96 frames per second, enabling online verification with imperceptible user delay and making \SystemName suitable for time-sensitive scenarios such as secure access and live deepfake defense.

\vspace{-0.1in}

\section{Discussion}
\label{sec:discussion}

\textbf{Resilience to Physical-Layer Spoofing.}
Unlike periodic signals such as heartbeats~\cite{oshim2025anti}, throat vibrations are aperiodic and phoneme-synchronized, making them significantly harder to spoof.
Stronger adaptive attacks---such as replay-and-mimicry, mechanical actuators, RF injection~\cite{li2024practical}, or coordinated audio+mmWave synthesis---would require generating fine-grained, physically consistent, tightly synchronized mmWave traces that match not only the audio, but also the underlying speech production process, device geometry, and RF channel constraints, which remains technically demanding.
Future work could incorporate radar challenge-response mechanisms to further strengthen defense.

\noindent\textbf{Defense against Unvoiced Phoneme Forgery.}
Although unvoiced phonemes (e.g., /s/, /t/) exhibit weak glottal vibration, \SystemName operates on short spectrogram windows capturing local phonetic context. Genuine unvoiced segments appear as structured low-amplitude regions within coherent articulatory patterns from adjacent voiced transitions, whereas replay attacks produce mechanical micro-Doppler artifacts (Fig.~\ref{fig:phySpe}) that serve as discriminative cues.

\noindent\textbf{Deployment Scenarios.}
\SystemName targets high-stakes, proximal capture-time verification, such as secure communication, media forensics, online oral examinations, and device authentication.
Users are typically stationary and within 50~cm, consistent with the system's effective range.
Although consumer devices do not yet support mmWave speech sensing, the trend toward mmWave-equipped smartphones (e.g., Google Pixel 4 and Samsung Galaxy S21 Ultra) and FMCW miniaturization suggests increasing deployment feasibility.

\noindent\textbf{Superiority over Reconstruction-based Approaches.}
Prior mmWave works reconstruct audible speech from vibration data~\cite{wang2022wavesdropper}, but reconstruction is ill-posed and discards fine-grained physical cues needed for spoof detection.
\SystemName instead models cross-modal consistency in a shared latent space, preserving micro-articulatory signatures with higher sensitivity to tampering.

\vspace{-0.05in}

\section{Related Work}
\label{sec:related}
\textbf{Audio Anti-Spoofing and Deepfake Detection.}
Audio anti-spoofing has evolved from handcrafted features~\cite{sahidullah2015comparison,javed2022voice} to deep learning~\cite{gomez2019light,chitale2024hybrid,tak2021end}, self-supervised~\cite{tak2022automatic,li2023voice}, and contrastive methods~\cite{wu2024clad}, but these degrade under splicing or editing~\cite{li2025measuring}.
Consistency-based approaches verify alignment with auxiliary modalities such as lip-speech synchronization~\cite{bohacek2024lost,Zhou_2021_ICCV} or signed QR codes~\cite{shahid2023my}, yet require visual input.
Physically grounded methods~\cite{kumari2025voiceradar,blue2022you,gao2025exploring,duan2024f2key,duan2024earse} reconstruct vocal-tract parameters~\cite{blue2022you} or model room impulse responses~\cite{kumari2025voiceradar}, but remain in the digital domain; Aegis~\cite{gao2025exploring} and F$^2$Key~\cite{duan2024f2key} add physical sensing yet require dedicated user-side hardware.
Our work instead passively senses throat vibrations via radar for physical consistency checks.

\noindent\textbf{mmWave-based Audio Sensing.}
mmWave has been used for eavesdropping via vocal fold vibrations~\cite{wang2022wavesdropper} and device-surface vibrations~\cite{basak2022mmspy,wang2022mmeve,feng2023mmeavesdropper,hu2022milliear,wang2022mmphone,basak2025mmwave}, as well as speech enhancement including keyword spotting~\cite{xu2019waveear,zhao2023radio2text}, separation~\cite{liu2021wavoice}, multi-modal fusion~\cite{fan2023mmmic}, and loudspeaker encoding~\cite{cui2024talk2radar}.
VocalPrint~\cite{li2020vocalprint} and WavoID~\cite{liu2023wavoid} use throat vibrations for speaker verification, and OriginStory~\cite{originstory2025asu} proposes grounding media authenticity in physiological signals.
While these works focus on speech recovery, interaction, or \emph{who} spoke, \SystemName addresses a distinct problem: verifying \emph{what} was spoken at word-level granularity, requiring tight segment-level alignment and fine-grained mismatch scoring to detect localized edits that preserve speaker identity.

\section{Conclusion}
\label{sec:conclusion}
This paper presents \SystemName, a novel system that leverages cross-modal physical consistency between mmWave and audio for fine-grained audio forgery detection. 
Unlike traditional methods that rely solely on digital acoustic statistics, \SystemName establishes a hard-to-forge physical anchor rooted in the physiological vocal production process.
We introduce an end-to-end framework featuring phase-aware vibration extraction, hierarchical cross-modal alignment, and a contrastive coherence network for word-level forgery detection.
This work establishes a new paradigm for physically grounded, cross-modal authentication, paving the way for trustworthy voice verification in forensic analysis, media integrity, and high-stakes security applications.

\newpage
\bibliographystyle{ACM-Reference-Format}
\bibliography{main}

\end{document}